\documentclass[pra,aps,showpacs,twocolumn,floatfix,superscriptaddress]{revtex4-1}
\usepackage{graphicx}
\usepackage[ansinew]{inputenc}
\usepackage{array}
\usepackage{color}
\usepackage{amsmath}
\usepackage{amsxtra}
\usepackage{amstext}
\usepackage{amssymb}
\usepackage{latexsym}
\usepackage{dsfont}
\begin{document}

\title{Theory of an optomechanical quantum heat engine}

\author{Keye Zhang}
\affiliation{Quantum Institute for Light and Atoms, Department of Physics, East
China Normal University, Shanghai, 200241, People's Republic of China }
\affiliation{B2 Institute, Department of Physics and College of Optical Sciences,
University of Arizona, Tucson, Arizona 85721, USA}
\author{Francesco Bariani}
\affiliation{B2 Institute, Department of Physics and College of Optical Sciences,
University of Arizona, Tucson, Arizona 85721, USA}
\author{Pierre Meystre}
\affiliation{B2 Institute, Department of Physics and College of Optical Sciences,
University of Arizona, Tucson, Arizona 85721, USA}

\begin{abstract}
Coherent interconversion between optical and mechanical excitations in an optomechanical cavity can be used to engineer a quantum heat engine. This heat engine is based on an Otto cycle between a cold photonic reservoir and a hot phononic reservoir [Phys. Rev. Lett. \textbf{112}, 150602 (2014)]. Building on our previous work,  we (i) develop a detailed theoretical analysis of the work and the efficiency of the engine, and (ii)  perform an investigation of the quantum thermodynamics underlying this scheme. In particular, we analyze the thermodynamic performance in both the dressed polariton picture and the original bare photon and phonon picture. Finally, (iii) a numerical simulation is performed to derive the full evolution of the quantum optomechanical system during the Otto cycle, by taking into account all relevant sources of noise.
\end{abstract}

\maketitle

\section{Introduction}

Optomechanical systems have witnessed spectacular developments in the last decade and can now operate deep in the quantum regime (see e.g.~Refs~\cite{RMP,Meystre,DSK} for recent reviews).  Conventional cryogenic cooling for mechanical oscillators of relatively high frequencies in the gigahertz range or higher~\cite{Oconnell}, and alternatively sideband cooling at lower mechanical frequencies~\cite{Teufel,Painter} have succeeded in bringing mechanical oscillators close to their quantum mechanical ground state. Also, quantum entanglement and squeezed states of photons and phonons have been demonstrated in these systems~\cite{Palomaki13,Naeini13}.

These developments pave the way to the creation of a generation of quantum interfaces between light and mechanical systems with broad potential for applications in quantum technology. One example is the coherent interconversion between optical and mechanical excitations, which was proposed and analyzed in a number of earlier theoretical studies \cite{interface}, and it has been used for the experimental realization of optomechanical light storage and readout \cite{interfaceEX}.  These effects are based on the mixing of photons and phonons into polariton normal modes. Importantly in the context of the present paper, these excitations are in contact with the thermal reservoirs of the cavity mode and the mechanics which may have a large temperature difference. It is then possible to envision a quantum heat engine whose ``working fluid'' is a polariton mode of the optomechanical system \cite{ourPRL}. In this heat engine the properties of the polariton are controlled by the cavity-pump detuning: by adiabatically switching between the phonon side and the photon side, and enabling thermalization with the corresponding reservoirs, we may realize a quantum Otto cycle. Such a heat engine working deep in the quantum regime may have the potential for challenging the classic law of thermodynamics \cite{Scullyengine, ionengine}.
Furthermore, consideration of the fast development in nano- and microelectromechanical systems (NEMS and MEMS) suggests that a quantum engine based on an optomechanical system may prove attractive in terms of manipulation, integration, and application.

This paper presents a detailed theoretical analysis of the work and the efficiency of the optomechanical heat engine and investigates the quantum thermodynamics involved. It compares the thermodynamics in the normal mode and in the bare mode pictures, the associated interpretations of the physics bringing to the fore subtle aspects of the role of quantum correlations.  It concludes by presenting results of numerical simulations of the full evolution of the Otto cycle, including dissipation and noise, with parameters within reach of existing technology.

The paper is organized as follows. Section II outlines the quantum model of the optomechanical system and  analyzes key features of the polariton modes for various values of the cavity-pump detuning. Section III describes the four stages of the Otto cycle and derives expressions for heat exchanged with reservoirs and work delivered. Section IV discusses the limit of the thermal efficiency of the engine at maximum work. Section V analyzes the effective master equation for the polaritons and discusses implications of the fact that they are coupled to squeezed baths when the bare modes of the system are coupled to thermal reservoirs. Section VI turns the thermodynamics of the whole system and of its subsystems. An intuitive physical picture of the optomechanical engine is also suggested.  Finally, Sec. VII presents selected results of full numerical simulations of the engine, and Sec. VIII is a summary and outlook. 

\section{The optomechanical system}
We consider a generic optomechanical system consisting of a Fabry-P{\'e}rot resonator with a compliant end mirror of effective mass $m$ and frequency $\omega_m$ driven by the radiation pressure from a single-mode intracavity field. We assume the system has reached a mean-field steady state characterized by a classical intracavity field $\alpha$ and corresponding normalized mirror displacement $x/x_{\rm zpt}=\beta$, where $x_{\rm zpt}=(\hbar/2m \omega_m)^{1/2}$ is the zero-point mirror displacement. For small optical damping rates $\kappa \ll |\Delta_p|$ we have $\alpha \approx \alpha_{\rm in}/\Delta_p$, where $\alpha_{\rm in}$ is the amplitude of the pump and $\beta \approx -g_0\alpha^2/\omega_m$. 

This system is described by the linearized optomechanical Hamiltonian  \cite{steady}
\begin{equation}
H_{0}/\hbar=-\Delta_{p}\hat a^{\dagger}\hat a+\omega_{m}\hat b^{\dagger}\hat b+G(\hat b+\hat b^{\dagger})(\hat a+\hat a^{\dagger}).
\label{eq:H0}
\end{equation}
Here $\hat a$ is the photon annihilation operator for the quantum fluctuations of the optical mode of frequency $\omega_c$ driven by a classical pump of frequency $\omega_p$,  $\hat b$ is the operator describing the quantum fluctuations of the mechanics,
\begin{equation}
\Delta_{p}=\omega_{p}-\omega_{c}-2g_{0}\beta,
\end{equation}
is the effective detuning between the optical pump and the cavity mode, and $g_0$ is the single-photon optomechanical coupling.  Finally the linearized effective optomechanical coupling is 
\begin{equation}
G=g_{0}\alpha.
\end{equation}
We take it to be real and positive in this work without loss of generality. 

\begin{figure}[]
\includegraphics[width=0.45 \textwidth]{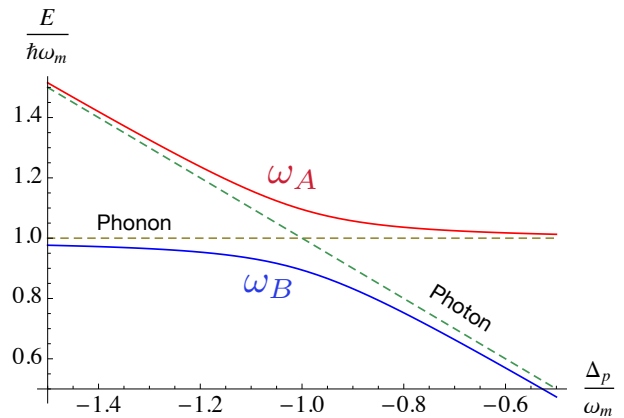}
\caption{Frequencies of the two polaritons (normal modes) of the optomechanical system for $G/\omega_{m}=0.05$ in the red-detuned case $\Delta_p < 0$. The dashed curves correspond to the frequencies of the bare photon and phonon modes.}
\label{fig:Two-branches}
\end{figure}

The Hamiltonian (\ref{eq:H0}) can be diagonalized in terms of two uncoupled bosonic normal modes, or polaritons, with annihilation operators $\hat A$ and $\hat B$ as
\begin{equation}
H_0=\hbar \omega_{A}\hat A^{\dagger}\hat A +\hbar \omega_{B}\hat B^{\dagger}\hat B + {\rm const.},\label{HAB}
\end{equation}
with corresponding eigenfrequencies
\begin{eqnarray}
\omega_{A} & = & \frac{1}{\sqrt{2}}\sqrt{\Delta_{p}^{2}+\omega_{m}^{2}+\sqrt{(\Delta_{p}^{2}-\omega_{m}^{2})^{2}-16G^{2}\Delta_{p}\omega_{m}}},\\
\omega_{B} & = & \frac{1}{\sqrt{2}}\sqrt{\Delta_{p}^{2}+\omega_{m}^{2}-\sqrt{(\Delta_{p}^{2}-\omega_{m}^{2})^{2}-16G^{2}\Delta_{p}\omega_{m}}}
\label{eq:omn}
\end{eqnarray}
(see Fig.~ \ref{fig:Two-branches}). 

We consider the red-detuned regime ($\Delta_p<0$) where the beam-splitter interaction term $G(\hat{a}^\dagger\hat{b}+\hat{b}^\dagger\hat{a})$ plays a dominant role and the stability condition of the linearized optomechanical system ($\omega_B>0$) gives
\begin{equation}
\Delta_p<-\frac{4G^2}{\omega_m}.\label{stable}
\end{equation}  
To  second order in $G/\omega_{m}$ and for $\Delta_{p}<-\omega_{m}$, the normal mode frequencies reduce to
\begin{eqnarray}
\omega_{A} & \approx&-\Delta_{p}\left (1-\frac{2G^{2}\omega_{m}}{(\Delta_{p}^{2}-\omega_{m}^{2})\Delta_{p}}\right ), \nonumber \\
\quad\omega_{B}&\approx&\omega_{m}\left (1+\frac{2\Delta_{p}G^{2}}{(\Delta_{p}^{2}-\omega_{m}^{2})\omega_{m}}\right ).
\end{eqnarray}
For $\Delta_{p}/\omega_{m}\rightarrow-\infty$ we have $\omega_{A} \rightarrow -\Delta_{p}$ and $\omega_{B} \rightarrow \omega_{m}$, so that $A$ describes a photonlike excitation and $B$  a phononlike excitation. In contrast, for  $-\omega_{m}<\Delta_{p}<0$ we have
\begin{eqnarray}
\omega_{A}&\approx&\omega_{m}\left (1+\frac{2\Delta_{p}G^{2}}{(\Delta_{p}^{2}-\omega_{m}^{2})\omega_{m}}\right ),\nonumber \\
\omega_{B} &\approx& -\Delta_{p}\left (1-\frac{2G^{2}\omega_{m}}{(\Delta_{p}^{2}-\omega_{m}^{2})\Delta_{p}}\right ).
\end{eqnarray}
The polariton $A$ is then phononlike while $B$ is photonlike as  $\Delta_{p}\rightarrow 0^{(-)}$. At the avoided crossing $\Delta_{p}=-\omega_{m}$, we have 
\begin{equation}
\omega_{A,B}=\omega_{m}\sqrt{1\pm\frac{2G}{\omega_{m}}},
\label{eq:gap}
\end{equation}
which shows that the minimum frequency difference between branches $A$, $B$ is proportional to $G/\omega_m$.

\section{Otto Cycles}
\label{sec:Otto-Cycle}

So far, we have only discussed the coherent contribution for the dynamics of the optomechanical system. Accounting in addition for optical and mechanical dissipation allows one to  exploit the two thermal reservoirs to engineer a heat engine working between the ``hot" thermal bath responsible for the relaxation of the phonon mode and the ``cold" thermal bath due to the damping of the optical mode. As discussed in Ref.~\cite{ourPRL}, it is then possible to operate the optomechanical system as a quantum Otto cycle \cite{QuOtto} by varying the detuning $\Delta_p$ while keeping the intracavity optical field $\alpha$ constant. Provided that nonadiabatic transitions between the two polariton branches can be avoided, each band can be associated with a different Otto cycle. We now turn to a detailed discussion of these cycles. 

\begin{figure}
\includegraphics[width=0.4\textwidth]{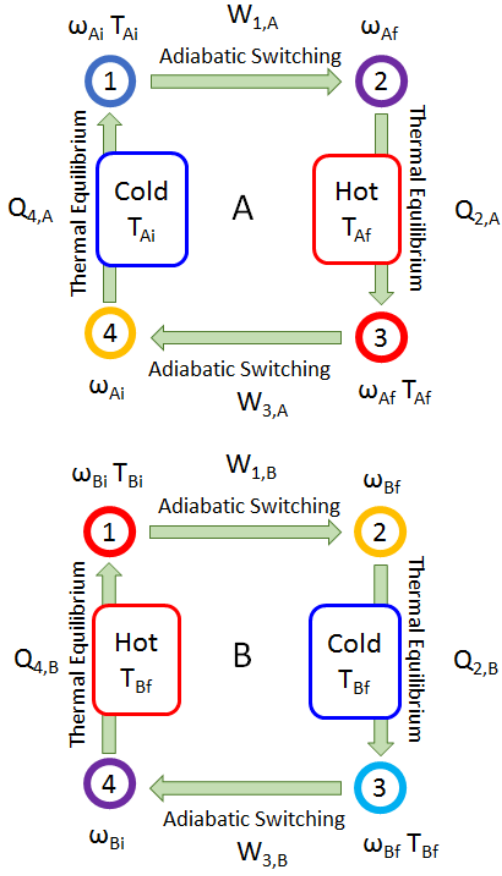}
\caption{Schematics of the Otto cycles associated with the polariton branches $A$ and $B$. See text for details.}
\label{fig:Otto-cycles}
\end{figure}

We consider a situation where the optomechanical system is initially in thermal equilibrium at large red detuning, $\Delta_{p}/\omega_{m}\sim-\infty$, so that the phononlike lower polariton branch $B$ is in thermal equilibrium with a reservoir at effective temperature $T_{Bi}$ --- for all practical purposes the temperature of the phonon heat reservoir. Similarly, the photonlike upper polariton branch $A$ is in thermal equilibrium with a reservoir at  temperature $T_{Ai} \approx 0\ \rm{K}$, an excellent approximation at optical frequencies.  Since $T_{Bi}\gg T_{Ai}$ we have the initial polariton population
\begin{equation}
\langle \hat{N}_{B}\rangle_i\gg\langle \hat{N}_{A}\rangle_i.\label{eq:BLA}
\end{equation}

The first stroke of the cycle is an adiabatic change of $\Delta_p$  from its initial value $\delta_i = \Delta_{pi}/\omega_m\sim-\infty$ to a final  value $\delta_f = \Delta_{pf}/\omega_{m}\rightarrow0^{(-)}$; this step has to be fast enough that the interaction of the system with the thermal reservoirs can be largely neglected,  yet slow enough that nonadiabatic transitions between the two polariton branches are negligible. Ideally, at the end of the stroke the lower-branch polariton becomes photonlike. It is then allowed to reach thermal equilibrium with a reservoir at temperature $T_{Bf} \approx 0\ \rm{K}$, the temperature of the photon reservoir, while the upper polariton branch relaxes to the temperature of the phonon reservoir. This is the second stroke. The third stroke of the cycle involves sweeping the detuning back to its initial large negative value. Again, this step has to be fast enough to avoid thermalization, but slow enough to avoid nonadiabatic transitions. The final stroke is the rethermalization at fixed detuning $\delta_i$: this leads to essentially the temperature of the phonon reservoir, $T_{Bi}$, for the lower polariton branch and to $T_{Ai}$ for the upper branch. We stress that the amplitude of the driving classical field needs to be adjusted during the detuning changes so that the intracavity amplitude is kept constant. The Otto cycles associated with the two polariton branches are sketched in Fig.~\ref{fig:Otto-cycles}.

Denoting by $E_{i,\alpha}$, $i= 1, \ldots, 4$ and $\alpha = \{A, B\}$ the energies of the system at the four nodes of these cycles, we have that the heat exchanged and the work performed during each stroke are given by
\begin{eqnarray}
W_{1,\alpha} & = & E_{2,\alpha}-E_{1, \alpha},\nonumber \\
Q_{2, \alpha} & = & E_{3,\alpha}-E_{2,\alpha},\nonumber \\
W_{3,\alpha} & = & E_{4,\alpha}-E_{3,\alpha},\nonumber \\
Q_{4,\alpha} & = & E_{1,\alpha}-E_{4,\alpha}.
\label{eq:QW}
\end{eqnarray}
with
\begin{equation}
W_{1,\alpha}+W_{3,\alpha}+Q_{2,\alpha}+Q_{4,\alpha}=0.
\end{equation}
According to the Hamiltonian (\ref{HAB}), $E_{i,\alpha}$ is dependent on the detuning and the expectation value of the polariton number, so the total work on the two polariton cycles is
\begin{equation}
W_{\alpha, \rm tot}  =  W_{1,\alpha}+W_{3,\alpha}=\hbar(\omega_{\alpha i}-\omega_{\alpha f})\left (\langle \hat{N}_{\alpha}\rangle_{f} -\langle \hat{N}_{\alpha}\rangle_{i}\right ).\label{eq:WB}
\end{equation}
Here $\omega_{\alpha l}$ and $\langle \hat{N}_{\alpha}\rangle_{l}$, with $l=\{i,f\}$, are the frequencies and thermal mean populations of the two polariton modes at the initial and the final detunings, respectively. Since $\omega_{Ai}-\omega_{Af}>0$ and $\langle \hat{N}_{A}\rangle_{f} -\langle \hat{N}_{A}\rangle_{i}>0$ we have that $W_{A, \rm tot}>0$. Similarly, $\omega_{Bi}-\omega_{Bf}>0$ and $\langle \hat{N}_{B}\rangle_{f} -\langle \hat{N}_{B}\rangle_{i} <0$ so that $W_{B, \rm tot}<0$. That is, in the thermal cycle operated along the polariton branch $A$ work is performed {\it on} the system, associated with the release of heat. In contrast, in the cycle along branch $B$ heat is absorbed by the system with corresponding work performed {\it by} the system. Cycle $B$ thus represents an Otto heat engine: it receives heat from the high temperature reservoir and partially converts it to work, while releasing the remaining heat to the low temperature reservoir. This is the process we are interested in. 

\begin{figure*}
\includegraphics[width=0.8 \textwidth]{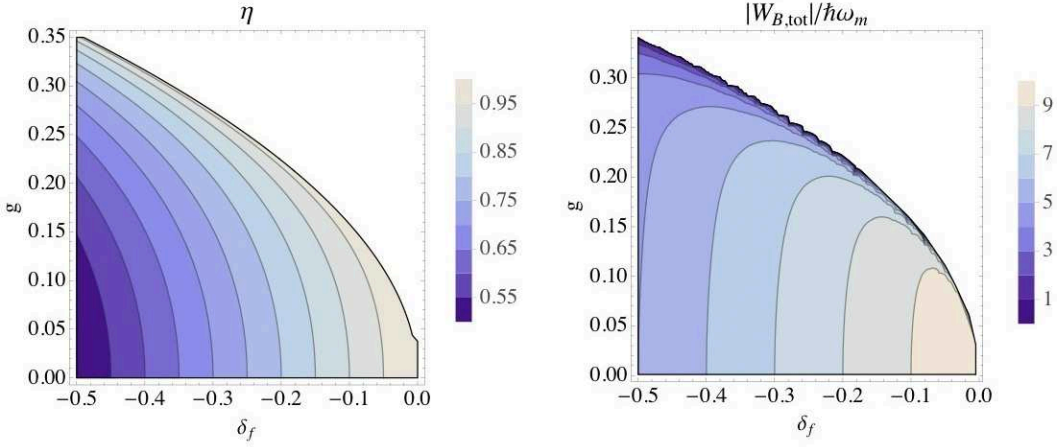}
\caption{(Color online) Contour maps of the thermal efficiency and the total work (absolute value) of the $B$-branch Otto cycle for $\delta_{i}=-3$ with value legends aside. The thermal mean population of the normal mode $B$ is calculated through a numerical Bogoliubov transformation with the thermal mean photon and phonon numbers, $\bar{n}_a=0$ and $\bar{n}_{b}=10$, respectively. The white region is mechanically unstable. 
\label{fig:The-thermal-efficient}}
\end{figure*}

The efficiency of cycle $B$ is defined by the ratio between the total work and the input heat \cite{ionengine}
\begin{equation}
\eta_{B}=\frac{-W_{B,\rm tot}}{Q_{4,B}}=1-\frac{\omega_{Bf}}{\omega_{Bi}}.
\label{eq:etaB}
\end{equation}
Figure~\ref{fig:The-thermal-efficient} shows $\eta_{B}$ and $W_{B, \rm tot}$ as a function of $\delta_{f}$ and the dimensionless interaction strength $g=G/\omega_m$. (We do not show their dependence on $\delta_i$ because the energy spectrum of mode $B$ is weakly dependent on it for large negative values.) The efficiency $\eta_B$ is independent of the thermal mean polariton number and is maximized for $\omega_{Bf}=0$, which is precisely the stability condition of the system, see Eqs.~(\ref{eq:omn}) and (\ref{stable}). This means that the thermal efficiency could be large even for large optomechanical interaction strengths and final detuning far from zero, provided that they are close to the instability threshold~(\ref{stable}), $\Delta_{pf}=-4G^2/\omega_m$.  However, this is not the case for the total work $W_{B, \rm tot}$ that reaches its maximum value at small $g$ and with a near-resonant $\delta_f\approx 0$.

We finally note that the cycle $A$ is a reversed engine whose efficiency is defined by the ratio between the total work and the output heat
\begin{equation}
\eta_{A}=\frac{W_{A,tot}}{-Q_{4,A}}=1-\frac{\omega_{Af}}{\omega_{Ai}}.
\end{equation}
In the limiting case $\delta_{i}\rightarrow -\infty$, $\delta_f \rightarrow 0^{(-)}$, and $g\rightarrow 0$, we find  $W_{A, \rm tot}+W_{B, \rm tot}\approx 0$. In order to maximize the work extracted from the system we should therefore avoid the occurrence of nonadiabatic transitions between the two cycles.

\section{Thermal efficiency}
We now proceed with a more quantitative description of the Otto cycle along branch $B$. For the situations considered here, where the polariton system is adiabatically switched from the phononlike side to the photonlike side and back, it is convenient to work in the bare mode picture, rather than with the dressed polariton modes. There are however potential issues with this approach. These are discussed in the following section.

\subsection{Normal modes and bare modes}
We proceed by expressing the annihilation and creation operators of the polaritons in terms of the bare modes via the Bogoliubov transformation
\begin{equation}
\begin{pmatrix}
\hat A\\
\hat B\\
\hat A^{\dagger}\\
\hat B^{\dagger}
\end{pmatrix}=\begin{pmatrix}U^{\dagger} & -V^{\dagger}\\
-V^{T} & U^{T}
\end{pmatrix}
\begin{pmatrix}
\hat a\\
\hat b\\
\hat a^{\dagger}\\
\hat b^{\dagger}
\end{pmatrix},\label{eq:transform}
\end{equation}
where $U$ and $V$ are $2\times2$ submatrices that satisfy the relationships
\begin{eqnarray}
U^{\dagger}U-V^{\dagger}V & = & I,\label{eq:condition1}\\
U^{\text{T}}V-V^{\text T}U & = & 0,\label{eq:condition2}
\end{eqnarray}
with the inverse transformation
\begin{equation}
\begin{pmatrix}
\hat a\\
\hat b\\
\hat a^{\dagger}\\
\hat b^{\dagger}
\end{pmatrix}=
\begin{pmatrix}U & V^{*}\\
V & U^{*}
\end{pmatrix}
\begin{pmatrix}
\hat A\\
\hat B\\
\hat A^{\dagger}\\
\hat B^{\dagger}
\end{pmatrix}.\label{eq:inverse transform}
\end{equation}

In the limit of small dimensionless optomechanical couplings $g \ll 1$ and for detunings $\delta \ll -1$ we find, to second order in $g$,
\begin{widetext}
\begin{eqnarray}
\hat A & = & \left[1+\frac{2\delta g^2}{(\delta-1)^2}\right]{\hat a}-\frac{g}{1+\delta}{\hat b} + \frac{g^2}{\delta(1-\delta^2)}{\hat a}^{\dagger} + \frac{g}{1-\delta}{\hat b^\dagger},\\
\hat B & = & \frac{g}{1+\delta}\hat a+\left[1+\frac{2\delta g^2}{(\delta-1)^2}\right]{\hat b} + \frac{g}{1-\delta}{\hat a}^{\dagger} + \frac{g^{2}\delta}{\delta^{2}-1}\hat b^{\dagger},
\end{eqnarray}
and
\begin{eqnarray}
\hat N_A=\hat A^{\dagger}\hat A & = & \left[1+\frac{4\delta g^{2}}{(\delta-1)^{2}}\right]\hat a^{\dagger}\hat a+\frac{2(1+\delta^{2})g^{2}}{(\delta-1)^{2}}\hat b^{\dagger}\hat b+\left(\frac{g}{1-\delta}\right)^{2} \nonumber\\
 &  & -\frac{g}{1+\delta}(\hat a^{\dagger}\hat b+\hat b^{\dagger}\hat a)+\frac{g^{2}}{\delta(1-\delta^{2})}(\hat a^{2}+\hat a^{\dagger2})+\frac{g}{1-\delta}(\hat a\hat b+\hat b^{\dagger}\hat a^{\dagger})-\frac{g^{2}}{1-\delta^{2}}(\hat b^{2}+\hat b^{\dagger2}),\label{NA}\\
\hat N_B=\hat B^{\dagger}\hat B & = & \left[1+\frac{4\delta g^{2}}{(\delta-1)^{2}}\right]\hat b^{\dagger}\hat b+\frac{2(1+\delta^{2})g^{2}}{(\delta-1)^{2}}\hat a^{\dagger}\hat a+\left(\frac{g}{1-\delta}\right)^{2} \nonumber\\
 &  & +\frac{g}{1+\delta}(\hat a^{\dagger}\hat b+\hat b^{\dagger}\hat a)+\frac{g^{2}}{1-\delta^{2}}(\hat a^{2}+\hat a^{\dagger2})+\frac{g}{1-\delta}(\hat a\hat b+\hat b^{\dagger}\hat a^{\dagger})+\frac{g^{2}\delta}{\delta^{2}-1}(\hat b^{2}+\hat b^{\dagger2}),\label{NB}
\end{eqnarray}
\end{widetext}
from which the steady-state mean population of the polariton modes can be expressed in terms of mean photon and phonon occupations, second-order photon-phonon correlations, and a term associated with squeezing. For small optomechanical coupling strengths $g$ and far from the sideband resonance at $\delta=-1$, we can neglect these correlations and squeezing, and furthermore approximate the mean photon and phonon numbers as the mean thermal occupations of the optical reservoir $\bar n_a$ ( $\bar n_a\approx 0$ for optical frequencies) and of the mechanical reservoir $\bar n_b$, respectively.  More precisely, the steady populations of the polariton modes are approximated  by the first lines of Eqs. (\ref{NA}) and (\ref{NB}). In the limiting case $g\rightarrow 0$ the populations of the polaritons $A$ and $B$ approach the thermal photon number and the thermal phonon number, respectively. For detunings $-1<\delta <0$, the expressions for the operators $\hat A$ and $\hat B$ are simply interchanged. 

Clearly these simplifications cease to hold for larger $g$ and near the sideband resonance, in which case optomechanical entanglement and optomechanical cooling effects can play a significant role. In particular,  quantum correlations between photons and phonons can significantly reduce the phonon number from $\bar n_b$, leaving the bare photon and phonon modes, as well as the polariton modes, out of their thermal equilibrium \cite{Genes2008, Genes2008b}.  We investigate these features in some detail in Sec. \ref{MEpolariton}. 

\subsection{Efficiency}
Using the approximate expressions of $\hat N_A$ and $\hat N_B$ it is straightforward to evaluate the efficiency of the heat engine based on the polariton mode $B$. We assume that $\langle\hat{A}^{\dagger}\hat{A}\rangle = 0$ and keep the constant term to second order in the dimensionless optomechanical coupling $g$: this term affects the energy value at each node but it has no influence on the total work and the efficiency of the Otto cycle. Assuming that adiabatic transitions between the two polariton branches can be ignored, the Hamiltonian evolution is governed solely by
\begin{equation}
H_{B}=\hbar\omega_{B}\left[\hat B^{\dagger}\hat B-\left (\frac{g}{\delta-1}\right )^{2}\right].\label{HB}
\end{equation}
As already discussed $\delta_i \ll -1$, so the lower polariton branch $B$ is initially essentially phononlike and in thermal equilibrium with the phonon-dominated reservoir at temperature $T_{Bi}$, with mean thermal excitation 
\begin{equation}
\langle \hat N_B\rangle _{i} = \left[1+\frac{4\delta_{i}g^{2}}{(\delta_{i}^{2}-1)^{2}}\right]\bar{n}_{b}+\left(\frac{g}{1-\delta_i}\right)^{2},\label{NB1}
\end{equation}
so that
\begin{eqnarray}
E_{1,B}&=& \hbar\omega_{B}\left[\langle \hat{N}_B \rangle _{i}-\left(\frac{g}{\delta_{i}-1}\right)^{2}\right] \nonumber\\
&=&\hbar\omega_{m}\left(1+\frac{2\delta_{i}g^{2}}{\delta_{i}^{2}-1}\right)\left[1+\frac{4\delta_{i}g^{2}}{(\delta_{i}^{2}-1)^{2}}\right]\bar{n}_{b}.
\end{eqnarray}
Adiabatically changing the detuning to the new value  $\delta_f$ with $-1<\delta_{f}<0$ the energy of the polariton mode $B$ then becomes
\begin{eqnarray}
&&E_{2,B}= \hbar\omega_{m}\left(\frac{2g^{2}}{\delta_{f}^{2}-1}-\delta_{f}\right) \\
&\times&\left[\left(1+\frac{4\delta_{i}g^{2}}{(\delta_{i}^{2}-1)^{2}}\right)\bar{n}{}_{b}+\left(\frac{g}{\delta_{i}-1}\right)^{2}-\left(\frac{g}{\delta_{f}-1}\right)^{2}\right],\nonumber
\end{eqnarray}
where the population remains unchanged. After the system reaches its new thermal equilibrium with the photon-dominated reservoir at temperature $T_{Bf}$ and mean thermal excitation
\begin{equation}
\langle \hat N_B\rangle _{f} = \frac{2(1+\delta_f^{2})g^{2}}{(\delta_f-1)^{2}}\bar n_b+\left(\frac{g}{1-\delta_f}\right)^{2}, \label{NB2}
\end{equation}
its energy becomes
\begin{eqnarray}
E_{3,B}=\hbar\omega_{m}\left(\frac{2g^{2}}{\delta_{f}^{2}-1}-\delta_{f}\right)\left[\frac{2(1+\delta_{f}^{2})g^{2}}{(\delta_{f}^{2}-1)^{2}}\right]\bar{n}{}_{b}.
\end{eqnarray}
At this point, the detuning is changed back to $\delta_i$ and the system adiabatically returns to its phononlike nature, but still keeping the population (\ref{NB2}), so that 
\begin{eqnarray}
E_{4,B}&=&\hbar\omega_{m}\left(1+\frac{2\delta_{i}g^{2}}{\delta_{i}^{2}-1}\right) \\
&\times&\left[\left(\frac{2(1+\delta_f^2)g^2}{(\delta_{f}^{2}-1)^{2}}\right)\bar{n}{}_{b}+\left(\frac{g}{\delta_{f}-1}\right)^{2}-\left(\frac{g}{\delta_{i}-1}\right)^2\right].\nonumber
\end{eqnarray}
Finally, after  thermalization with the phonon-dominated bath, the energy returns to its initial value $E_{1,B}$. Combined with Eqs. (\ref{eq:QW}), (\ref{eq:WB}), and (\ref{eq:etaB}) this allows to determine the efficiency and total work of the cycle.

We first consider the limiting case  $g\rightarrow 0$. In this case the adiabaticity condition requires an infinite amount of time for the change in detuning to avoid the coupling of the two polariton branches, a condition in conflict with the requirement that thermalization remains insignificant during that stroke. Nonetheless this limit  provides useful insights into the physics of the system. We now have $\omega_{Bi}\approx\omega_{m}$, $\omega_{Bf}\approx-\Delta_{pf}$, $\langle \hat N_B\rangle _{i}\approx \bar{n}_b $ and $\langle \hat{N}_B\rangle _{f}\approx \bar{n}_a$.  Taking then the effective temperature of the photon reservoir to be $0$ K yields for the total work and efficiency (remember, $\delta_f=\Delta_{pf}/\omega_m<0$)
\begin{eqnarray}
W_{B,\rm tot} & = & -\hbar\omega_m(1+\delta_f)\bar{n}_b,\\
\eta & = & 1+\delta_f.
\end{eqnarray}
If we further assume $\delta_f\rightarrow0^{(-)}$, also an unrealistic situation, we then find that the thermal energy of the phonon can be fully converted into work. 

A more realistic estimate, consistent with the requirement to change the detuning $\delta$ adiabatically, can be obtained by evaluating these quantities to second order in $g$. Again, we take $\delta_{i}$ to be large and negative, and $\delta_{f}$ to be a small negative detuning close to zero, so that $\omega_{Bi}\approx \omega_{m}$ and $\omega_{Bf} \approx \omega_{m}(-\delta_{f}-2g^{2})$;
the thermal efficiency of the Otto cycle is then
\begin{equation}
\eta=1-(-\delta_{f}-2g^{2}),
\end{equation}
which is a maximum for $g^2=-\delta_f/2$. However, the total work
\begin{equation}
W_{B, \rm tot}=\hbar\omega_{m}(-\delta_{f}-2g^{2}-1)[(1-2g^{2})\bar{n}_b-g^{2}]
\end{equation}
reaches its minimum (remember, $W_{B, \rm tot}<0$) for 
\begin{equation}
g^{2}=-\frac{\delta_{f}}{4}-\frac{\hbar\omega_{m}}{8k_{B}T_{b}},\label{g2}
\end{equation}
where we have assumed a phonon temperature $T_{b}$ high enough that
\begin{equation}
\bar{n}_b\approx\frac{k_{B}T_{b}}{\hbar\omega_{m}}-\frac{1}{2}.
\end{equation}
This yields the efficiency at maximum power
\begin{equation}
\eta_{P}=1-\left (\frac{-\Delta_{pf}}{2\omega_{m}}+\frac{\hbar\omega_{m}}{4k_{B}T_{b}}\right ),
\end{equation}
which, with the help of a simple inequality, gives
\begin{equation}
\eta_{P}<1-\sqrt{\frac{\hbar(-\Delta_{pf})}{2k_{B}T_{b}}}.
\end{equation}
With a quantum-classical energy correspondence for the zero point energy of the cavity mode in the frame rotating at the cavity pump frequency, $-\hbar\Delta_{pf}/2 \sim k_BT_a$, we obtain the quantum version of the classical Curzon-Ahlborn efficiency limit, $1-\sqrt{T_{\rm low}/T_{\rm high}}$ \cite{CAlimit}. Its upper limit is reached for $-\Delta_{pf}/2\omega_m=\hbar\omega_m/(4k_{B}T_b)$
which, according to Eq.~(\ref{g2}),  corresponds to the ideal situation $g=0$.

\section{Master equation for the polariton}\label{MEpolariton}

When the optomechanical coupling is small but finite, all terms in Eqs.~(\ref{NA}) and (\ref{NB}) contribute, and the steady-state polariton $B$ occupation will deviate from thermal equilibrium. To investigate this effect we derive the effective master equation for the normal mode $B$ below. 

For the high-$Q$ mechanical oscillator that we consider it is safe to use the familiar Lindblad superoperator to describe the effect of  Brownian thermal motion on the mechanics~\cite{Mari2012}. In the bare mode picture, the master equation of the system is then
\begin{eqnarray}
\frac{d\rho}{dt}&=&-\frac{i}{\hbar}[H_0,\rho]+\kappa(\bar{n}_{a}+1)\mathcal{L}[\hat{a}]\rho+\kappa\bar{n}_{a}\mathcal{L}[\hat{a}^{\dagger}]\rho \nonumber\\
& &+\gamma(\bar{n}_{b}+1)\mathcal{L}[\hat{b}]\rho+ \gamma\bar{n}_{b}\mathcal{L}[\hat{b}^{\dagger}]\rho, \label{eq:master}
\end{eqnarray}
where $\bar{n}_{a}$ and $\bar n_b$ are the mean photon and phonon numbers in their respective thermal reservoirs, $\kappa$ and $\gamma$ are their decay rates, 
\begin{equation}
\mathcal{L}[\hat{x}]\rho=\hat{x}\rho \hat{x}^{\dagger}-\frac{1}{2}\hat{x}^{\dagger}\hat{x}\rho-\frac{1}{2}\rho \hat{x}^{\dagger}\hat{x},
\end{equation}
and $H_0$ is the linearized optomechanical Hamiltonian, Eq.~(\ref{eq:H0}). 

$H_0$ can be diagonalized via the Bogoliubov transformation~(\ref{eq:inverse transform});  however, the two polariton modes remain coupled via the Lindblad superoperators and it is not possible to define two uncoupled master equations for the normal modes $A$ and $B$. In the following we assume for simplicity that the population of the normal mode $A$ vanishes throughout the Otto cycle and we approximate the density matrix of the full system as
\begin{equation}
 \rho=\rho_{AB}\approx\rho_{B}\otimes\left|0\right\rangle \left\langle 0\right|_{A}.
 \end{equation}
In this case it is possible to obtain an effective master equation for the normal mode $B$ only,
\begin{eqnarray}
\frac{d\rho_{B}}{dt}&=&-\frac{i}{\hbar}[H_{B},\rho_{B}]\nonumber\\
& &+\Gamma_{B}(\bar{N}_{B}+1)\mathcal{L}[\hat B]\rho_{B}+\Gamma_{B}\bar{N}_{B}\mathcal{L}[\hat B^{\dagger}]\rho_{B} \nonumber\\
& &+\Gamma_{B}\bar{M}_{B}\mathcal{J}[\hat B]\rho_{B}+\Gamma_{B}\bar{M}_{B}^{*}\mathcal{J}[\hat B^{\dagger}]\rho_{B}. \label{eq:masterB}
\end{eqnarray}
where we have introduced the new superoperator
\begin{equation}
\mathit{\mathcal{J}}[\hat x]\rho=\hat x\rho \hat x-\frac{1}{2}\hat x\hat x\rho-\frac{1}{2}\rho \hat x\hat x,
\end{equation}
and the effective decay rate of the normal mode $B$
\begin{equation}
\Gamma_{B}=\kappa(|U_{12}|^{2}-|V_{12}|^{2})+\gamma(|U_{22}|^{2}-|V_{22}|^{2}).
\end{equation}
Here 
\begin{widetext}
\begin{eqnarray}
\bar{N}_{B}&=&\frac{\kappa(\bar{n}_{a}+1)\left|V_{12}\right|^{2}+\kappa\bar{n}_{a}\left|U_{12}\right|^{2}+\gamma(\bar{n}_{b}+1)\left|V_{22}\right|^{2}+\gamma\bar{n}_{b}\left|U_{22}\right|^{2}}{\kappa(|U_{12}|^{2}-|V_{12}|^{2})+\gamma(|U_{22}|^{2}-|V_{22}|^{2})},\\
\bar{M}_{B}&=&\frac{\kappa(2\bar{n}_{a}+1)V_{12}U_{12}+\gamma(2\bar{n}_{b}+1)V_{22}U_{22}}{\kappa(|U_{12}|^{2}-|V_{12}|^{2})+\gamma(|U_{22}|^{2}-|V_{22}|^{2})},
\end{eqnarray}
\end{widetext}
where $U_{ij}$ and $V_{ij}$ are the elements of the submatrices $U$ and $V$ of the Bogoliubov transformation.

The form of master equation (\ref{eq:masterB}) reveals that polariton $B$ is actually coupled to a squeezed thermal reservoir~\cite{Gardiner1992}. To characterize it we introduce the quadrature operators
\begin{eqnarray}
\hat X&=&\frac{1}{\sqrt{2}}(\hat B e^{i\omega_{B}t}+\hat B^{\dagger}e^{-i\omega_{B}t}),\\
\hat Y&=&\frac{1}{i\sqrt{2}}(\hat B e^{i\omega_{B}t}-\hat B^{\dagger}e^{-i\omega_{B}t}),
\end{eqnarray}
whose steady-state expectation values are easily found from the master equation~(\ref{eq:masterB}),
\begin{equation}
\langle \hat X\rangle _{s}=\langle \hat Y\rangle _{s}=0.
\end{equation}
with variances 
\begin{eqnarray}
\langle \hat X^{2}\rangle _{s} =\bar{N}_{B}-\bar{M}_{B}+\frac{1}{2},\\
\langle \hat Y^{2}\rangle _{s} =\bar{N}_{B}+\bar{M}_{B}+\frac{1}{2},
\end{eqnarray}
familiar from squeezed reservoirs. From the uncertainty relation
\begin{equation}
\left\langle X^{2}\right\rangle _{s}\left\langle Y^{2}\right\rangle _{s}\geqslant\frac{1}{4},
\end{equation}
we also find ($\bar M_B$ is taken to be real for simplicity in the following )
\begin{equation}
\bar{M}_{B}^{2}\leqslant\bar{N}_{B}(\bar{N}_{B}+1),
\end{equation}
with maximum squeezing reached for the equal sign. 

The presence of a squeezed reservoir implies that  the steady state of mode $B$ is not  a thermal state. Rather, it is a state that is in some sense ``hotter'' than the corresponding thermal reservoir. Its steady state population is
\begin{equation}
\bar{N}_{B} = \langle \hat B^{\dagger}\hat B\rangle _{s}=\frac{\langle \hat X^2\rangle_s+\langle \hat Y^2\rangle_s-1}{2},
\end{equation}
which is larger than the mean thermal population: For a squeezing parameter $r$ the general relationships between steady population $\bar N_B$ and thermal population $N_{\rm th}$ are
\cite{Marian1993, Breuer2007}
\begin{eqnarray}
\bar{N}_{B} & = & N_{\text{th}}+(2N_{\text{th}}+1)\sinh^{2}(r),\label{eq:sqthermal1}\\
\bar{M}_{B} & = & -\cosh(r)\sinh(r)(2N_{\text{th}}+1).\label{eq:sqthermal2}
\end{eqnarray}
This property was recently exploited in the ion heat engine scheme of Ref.~ \cite{ionengine2}, where the use of a squeezed reservoir was proposed to reach an efficiency that violates the familiar Carnot limit.  In our case both the cold reservoir (photonlike side) and the hot reservoir (phononlik side) are squeezed and due to the small coupling strength $g$ the squeezing effect is also very weak, making it a challenge to break the Carnot limit.

The exact expressions for the steady population $\bar N_B$ and effective decay rate $\Gamma_B$ are too cumbersome to be reproduced here. We give instead  their approximate forms to second order in $g$. For the case $\delta_i<-1$ and $\bar n_a=0$ we find
\begin{eqnarray}
\bar N_B &=& \left[1+\frac{4\delta_i g^{2}\kappa}{\gamma(\delta_{i}^{2}-1)^{2}}\right]\bar{n}_{b}+\frac{\kappa}{\gamma}\left(\frac{g}{1-\delta_i}\right)^{2},\label{NB3}\\
\Gamma_{B}&=&\gamma+(\gamma-\kappa)\frac{4g^{2}\delta_i}{(\delta_i^{2}-1)^{2}},
\end{eqnarray}
which show that not only the steady population but also the effective decay rate of the polariton $B$ are close to the phonon case for $\delta_i\rightarrow-\infty$.
On the other side where $-1<\delta_f<0$, the expressions become
\begin{eqnarray}
\bar N_B &=& \frac{2\gamma(1+\delta_f^{2})g^{2}}{\kappa(\delta_f-1)^{2}}\bar n_b+\frac{\gamma}{\kappa}\left(\frac{g}{1-\delta_f}\right)^{2},\label{NB4}\\
\Gamma_{B}&=&\kappa-(\gamma-\kappa)\frac{4g^{2}\delta_f}{(\delta_f^{2}-1)^{2}},
\end{eqnarray}
which tend to the photon case as $\delta_f\rightarrow0^{(-)}$. 

As a final note,  a comparison of Eqs.~(\ref{NB3}) and (\ref{NB1})  shows the important role of the ratio $\kappa/\gamma$ between the decay rates of the photons and the mechanics: In general, the terms in the second line of Eq.~(\ref{NB}) result in steady-state polariton populations that deviate from the thermal equilibrium result (\ref{NB1}). Neglecting the build up of correlations between the optical and phonon modes and of squeezing effects is strictly valid only for values of $\kappa/\gamma$ close to unity, corresponding to equal decay rates of the photon and the phonon. In practice, though, we found that, even for $\kappa/\gamma\gg1$, these effects are weak due to the assumption of small optomechanical coupling strength $g$. A similar situation occurs on the photonlike side, but with a reversed factor $\gamma/\kappa$; compare Eqs.~(\ref{NB4}) and (\ref{NB2}). 

\section{Quantum thermodynamics analysis}

So far, our discussion of the Otto cycle has been based on the polariton modes. This representation or, more precisely, the energy representation of the whole system is naturally required for the study of its thermodynamical properties. However, the thermodynamics of the subsystems, in this case, the photon and the phonon modes, is also of interest as it provides a more direct intuitive understanding of the underlying physics at play. With this in mind this section compares and contrasts the thermodynamics of the heat engine in the polariton and the bare mode pictures. 

\subsection{Work and heat exchange}
In classical thermodynamics, the expression of the first law is
\begin{equation}
dU=dQ+dW,
\end{equation}
where $U$, $Q$, and $W$, are energy, heat, and work, respectively. This law states that the energy exchanged by a system in a transformation is divided between work $W$ and heat $Q$. To obtain a quantum version of this expression, we express the average energy in terms of the eigenstates of the Hamiltonian $H$ as
\begin{equation}
U=\left\langle H\right\rangle =\sum_{i}p_{i}E_{i},
\end{equation}
where $E_{i}$ is the energy of the eigenstate $i$ with corresponding occupation probability $p_{i}$. An infinitesimal change in energy is then given by
\begin{equation}
dU=\sum_{i}dp_{i}E_{i}+\sum_{i}p_{i}dE_{i},
\end{equation}
and one can identify the first term on the right-hand side as the infinitesimal heat transferred, and the second as the infinitesimal work performed~\cite{firstlaw},
\begin{eqnarray}
dQ & = & \sum_{i}dp_{i}E_{i},\\
dW & = & \sum_{i}p_{i}dE_{i}.
\end{eqnarray}
The heat transferred to or from a quantum system corresponds to a change in the populations $p_i$ without change of the energy eigenvalues, while the work done on or by a quantum system corresponds to a redistribution of the energy eigenvalues. These quantum expressions of the infinitesimal heat and work are consistent with their definitions in classical thermodynamics and statistical physics. That is, the heat exchange results in a change in the statistical distribution of the microstates of different energies while the work is a change in the energy structure of the system.

One can also obtain more general expressions for $Q$ and $W$ in terms of the density operator $\rho(t)$ and the time-dependent Hamiltonian $H(t)$:
\begin{equation}
U(t)={\rm Tr}[\rho(t)H(t)].
\end{equation}
If we take the temporal derivative 
\begin{equation}
\partial_{t}U(t)={\rm Tr}[\partial_{t}\rho(t)H(t)]+{\rm Tr}[\rho(t)\partial_{t}H(t)],
\end{equation}
with
\begin{eqnarray}
\partial_{t}Q & = & {\rm Tr}[\partial_{t}\rho(t)H(t)],\\
\partial_{t}W & = & {\rm Tr}[\rho(t)\partial_{t}H(t)],\label{eq:dw}
\end{eqnarray}
then the general quantum definitions of $Q$ and $W$ are 
\begin{eqnarray}
Q & = & \int_{\rm cycle}{\rm Tr}[\partial_{t}\rho(t)H(t)]dt, \label{eq:Q}\\
W & = & \int_{\rm cycle}{\rm Tr}[\rho(t)\partial_{t}H(t)]dt.
\label{eq:W}
\end{eqnarray}

Let us then consider the first stroke of the optomechanical  Otto cycle, with the detuning $\delta$ adiabatically changed from a large negative value to a value close to zero so that the nature of the $B$ polariton changes from  phononlike to photonlike and the system outputs work. When considered in the polariton picture the adiabatic evolution ensures that the stroke is an isentropic process. But that interpretation only holds in the normal mode picture: While thermodynamical adiabaticity does mean that the system as a whole has no heat exchange with the environment,  heat can of course be exchanged between its subsystems, resulting in a change in the populations of their energy levels.

\begin{figure}[ptbh]
\includegraphics[width=0.4 \textwidth]{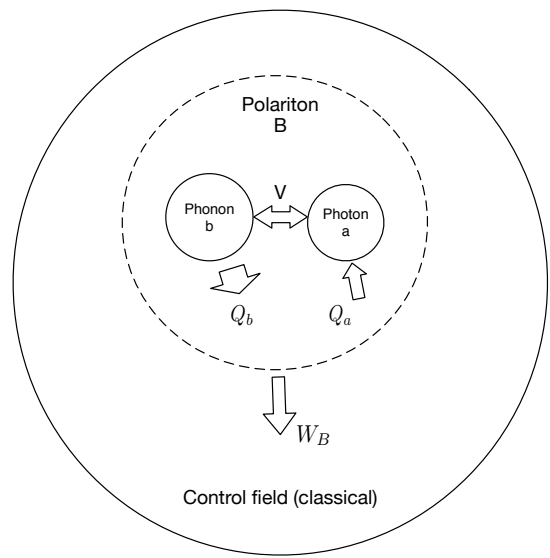}
\caption{Hierarchical structure of the quantum engine where the coupled photon and phonon modes constitute a polariton mode which exchanges heat and work with the external control field.}
\label{fig:The-hierarchical-diagram}
\end{figure}

To show how this works, we analyze the hierarchical structure of our system. As sketched in Fig.~\ref{fig:The-hierarchical-diagram} the first level is a bare mode picture, described by the Hamiltonian $H_0$ of Eq.~(\ref{eq:H0}) with the photon and phonon modes coupled by the linearized optomechanical interaction
\begin{equation}
V=G(\hat b+\hat b^{\dagger})(\hat a+\hat a^{\dagger}) \label{V}.
\end{equation}

The second level is the dressed picture, where the system is described in terms of the noninteracting normal modes (polaritons) $A$ and $B$. Here we ignore the polariton $A$, whose population remains negligible throughout the cycle, so that for all practical purposes the system is then described by the Hamiltonian $H_B$ (\ref{HB}). 

The third level, finally, includes the external controls. In our case they are the driving optical field, the steady cavity field $\alpha$, and the normalized displacement $\beta$. The temperatures of the photon and the phonon reservoirs should also be present at this level, but we ignore them during the isentropic stroke.

In both the bare modes and polariton pictures the change in average energy $U$ of the system is of course the same,
\begin{equation}
dU=d\left\langle H_{B}\right\rangle =d\left\langle H_{0}\right\rangle \label{eq:dU},
\end{equation}
but the interpretation of the thermodynamics is different. Specifically, in the polariton picture we have
\begin{equation}
d\left\langle H_{B}\right\rangle ={\rm Tr}[d\rho_{B}H_{B}]+{\rm Tr}[\rho_{B}dH_{B}],
\end{equation}
where $\rho_{B}$ is the density matrix of the normal mode $B$. Since the transformation is adiabatic, we have $d\rho_{B}=0$ so that 
\begin{eqnarray}
dQ_{B} & = & {\rm Tr}[d\rho_{B}H_{B}]=0\\
dU & = & {\rm Tr}[\rho_{B}dH_{B}]=dW_{B}.\label{dUWB}
\end{eqnarray}
Moreover, as the detuning $\Delta$ is changed from a large negative value to zero, $\omega_{B}$ decreases, so that  $dW_{B}<0$, indicative of the fact that work is produced by the heat engine. 

In contrast, in the bare picture we have
\begin{eqnarray}
d\left\langle H_{0}\right\rangle  & = & d\left\langle H_{a}\right\rangle +d\left\langle H_{b}\right\rangle +d\left\langle V\right\rangle \nonumber \\
 & = & {\rm Tr}[d\rho_{a}H_{a}]+{\rm Tr}[\rho_{a}dH_{a}]+{\rm Tr}[d\rho_{b}H_{b}]\nonumber \\
 &+&{\rm Tr}[\rho_{b}dH_{b}]+{\rm Tr}[d\rho_{ab}V]+{\rm Tr}[\rho_{ab}dV],
\end{eqnarray}
where $\rho_{ab}$ is the density matrix of the two-mode system and $\rho_a$ and $\rho_b$ are the reduced density matrices of the photon and phonon mode, respectively. Since $H_{b}$ and $V$ are constant, according to the quantum definitions of work and heat, we find
\begin{equation}
d\left\langle H_{0}\right\rangle =dQ_{a}+dW_{a}+dQ_{b}+{\rm Tr}[d\rho_{ab}V].\label{eq:dH0}
\end{equation}
By considering the change of the populations of the photon and phonon modes in the first stroke, we have 
\begin{eqnarray}
dQ_{a} & = & {\rm Tr}[d\rho_{a}H_{a}]>0,\\
dQ_{b} & = & {\rm Tr}[d\rho_{b}H_{b}]<0,
\end{eqnarray}
indicating that, in the bare picture, the evolutions of the photonic and phononic subsystems are neither adiabatic nor isentropic. Furthermore, since the initial population of the photon mode is zero, we have 
\begin{equation}
dW_{a}={\rm Tr}[\rho_{a}dH_{a}]=0.
\end{equation}
Finally, from Eqs.~(\ref{eq:dU}), (\ref{dUWB}), and (\ref{eq:dH0}) we find
\begin{equation}
dW_{B}=dQ_{a}+dQ_{b}+{\rm Tr}[d\rho_{ab}V], \label{eq:dWB}
\end{equation}
where the last term is the change of the quantum correlations between the photon and phonon fields: This is initially zero for a product of thermal states, but becomes finite as a result of the optomechanical coupling. This term is much smaller than $dQ_a$ and $dQ_b$ for the weak optomechanical couplings considered here, and interestingly, does not have a corresponding classical thermodynamical quantity.

Summarizing, in the dressed picture the first stroke of the heat engine adiabatically switches the polariton from a phononlike to a photonlike excitation and it performs work on the external control field. In the bare mode picture, the phonon mode releases heat, part of which is then absorbed by the photon field, a small amount contributing to quantum correlation, and the rest being absorbed by the external control field. Similar results can also be obtained for the second stroke of the Otto cycle, with $dW_B$ in Eq. (\ref{eq:dWB}) replaced by $dQ_B$, and $dQ_{a}<0$, $dQ_b>0$, corresponding to a process dominated by photon exothermic reaction (dissipation).  We stress that from Eqs. (\ref{eq:Q}) and (\ref{eq:W}), thermodynamics may be defined for the eigenstates of the system, namely the polaritons, while the application to the bare modes is only qualitative.

\subsection{Physical picture}
\begin{figure}
\includegraphics[scale=0.4]{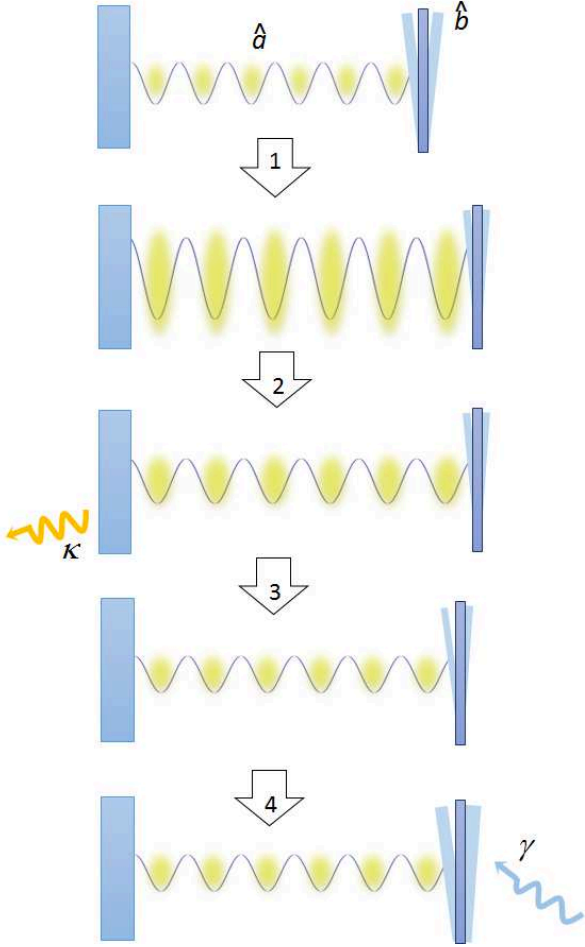}
\caption{Intuitive physical picture of the Otto cycle for the optomechanical heat engine. The numbered arrows  correspond to the engine strokes in Fig.~\ref{fig:Otto-cycles}. Initially (first figure) the mechanics undergoes relatively large thermal fluctuations due to its coupling to a hot thermal reservoir. After the adiabatic step 1, the polariton becomes photonlike, with its still-unchanged mean occupation becoming photonlike, resulting in added radiation pressure force on the mechanics. Thermalization at the low radiation field temperature significantly reduces the polariton occupation number in step 2. After the adiabatic step 3 the polariton has regained its phononic nature with a small adjustment of the mirror position. Finally the polariton is in contact with the hot thermal bath in step 4, regaining the initial thermal occupation number at rate $\gamma$. }
\label{fig:cycle}
\end{figure}

Following these considerations, one can gain a simple physical understanding of the engine cycle. As shown in Fig.~\ref{fig:cycle}, the change in cavity length (corresponding to the value of steady amplitude of the phonon field, $\beta$) represents the effect of the work performed by the engine. The vibration amplitudes of the cavity mirror and the cavity field represent the population of the phonon and photon mode, respectively.  

During stroke 1 the detuning  $\Delta$ is varied so that $\omega_p$  is brought closer to resonance with the cavity mode frequency $\omega_c$, while simultaneously changing the pumping rate $\alpha_{\rm in}$ so as to keep  mean intracavity amplitudes $\alpha$ and $\beta$, and hence the coupling frequency $G$, constant. As this happens the phononlike thermal excitations, which are initially large due to the contact with a thermal reservoir that is essentially at the temperature of the mechanics, are transformed into photonlike excitations. This occurs at a rate characterized by the coupling frequency $G$. During this step the amplitude of vibrations of the mechanics decreases and the excess energy is transferred to the intracavity field. As a result the resonator length increases slightly due to the increased radiation pressure. It is at this point that the mechanical work on the oscillator is produced by the optomechanical heat engine. However, this work is very small due to the disproportion between the steady amplitudes and the quantum fluctuations of the photon and phonon fields. During the thermalization step of stroke 2 the population of the photonlike excitations decays at rate $\kappa$ (for a photon reservoir at zero temperature) with the cavity length unchanged. In stroke 3 the remaining photonlike polariton excitations are turned back into phonon-like quanta by adjusting $\Delta$. This costs a small amount of work, resulting in a small contraction of the cavity length. The population of the phonon-like polariton excitation finally grows up to its initial value via thermal contact with the hot mechanical reservoir during stroke 4. The small polariton number in stroke (3) ensures that the total output work of the Otto cycle is positive. 
%\end{widetext}

\section{Numerical simulations}

\subsection{Time scales}
The adiabatic strokes 1 and 3 of the Otto cycle involve changes of the cavity detuning from $\delta_{i}\ll-1$ to $\delta_{f}\sim0^{(-)}$. To ensure the adiabaticity of the transformation, their times $\tau_1$ and $\tau_3$ must be much longer than the characteristic time of the transition between the two polariton branches $\delta\omega_{AB}^{-1}$.  From Eq.~(\ref{eq:gap}) we have
\begin{equation}
\tau_1,\tau_3\gg\frac{1}{\omega_{m}\sqrt{1+\frac{2G}{\omega_{m}}}-\omega_{m}\sqrt{1-\frac{2G}{\omega_{m}}}} \approx \frac{1}{2G},
\end{equation}
where we used the weak coupling condition $G/\omega_m\ll1.$ 
Additionally, in order to  avoid heat exchange between the normal modes and the reservoirs during the adiabatic strokes we also need  $\tau_1$ much shorter than the characteristic interaction time between the system and the reservoirs,
\begin{equation}
\tau_1\ll1/\kappa,1/\gamma.
\end{equation}
In addition, while for the detuning $\delta_{i}$ we have $\langle \hat N_{B}\rangle_i \gg \langle \hat N_{A}\rangle_i$ and can therefore safely neglect the upper polariton branch $A$, more care is needed in the second stroke, where $\langle \hat N_{A}\rangle_f\gg \langle \hat N_{B}\rangle_f$ at $\delta_{f}$.  To guarantee that the lower polariton branch reaches its photonic thermal equilibrium without significant increase in the population of the upper polariton branch requires therefore that 
\begin{equation}
\kappa>1/\tau_{2}\gg\gamma.
\end{equation}
Finally, the last stroke should be long enough for the polariton to rethermalize with the phonon reservoir. Combining these considerations results in the hierarchy of rates
\begin{equation}
1/\tau_{4}  <  \gamma  \ll  1/\tau_{2}  <  \kappa <  1/\tau_{1,3} \ll  G \ll\omega_{m}.\label{eq:hierarchy}
\end{equation}
The limiting factor for a fast execution of the cycle is then given by the slow thermalization with the mechanics, $\gamma$.  In addition, we assume in the simulations described in the following subsections that the optomechanical coupling $G$ is constant during the adiabatic strokes, a condition that can be satisfied by controlling the amplitude of the classical driving field as the cavity detuning is varied.

\subsection{Simulation results}

\begin{figure*}
\includegraphics[width=1 \textwidth]{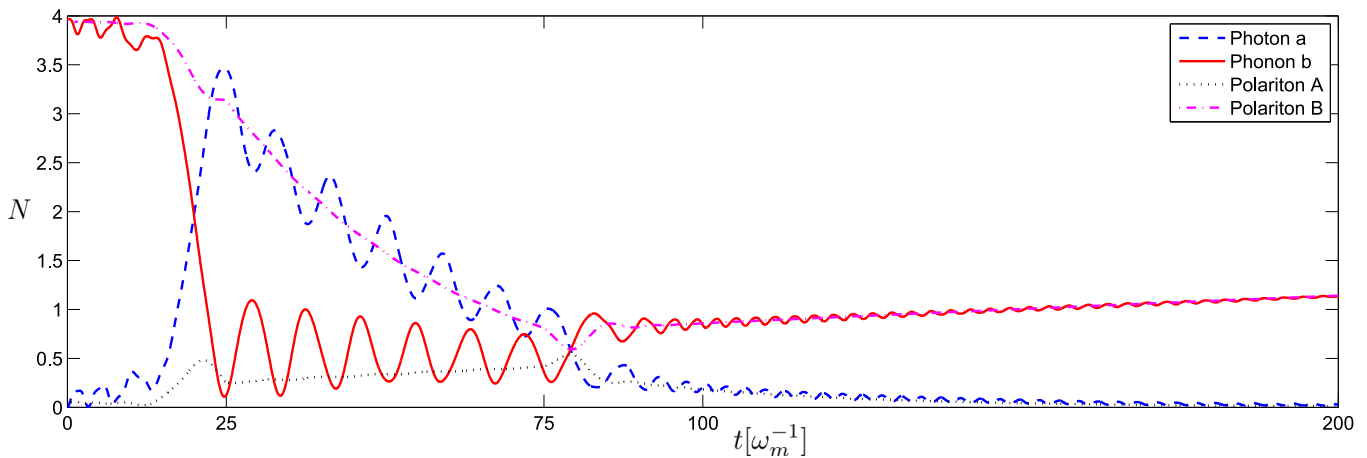}
\caption{(Color online) The time evolution of the number of the bare photon mode $a$ (blue dashed line), the bare phonon mode $b$ (red solid line), the polariton mode $A$ (black dotted line), and the polariton mode $B$ (pink dot-dashed line) in a loop of the Otto cycle. All parameters are normalized by the mechanical resonator frequency $\omega_m=2\pi\times200\ \rm{MHz}$. The optomechanical coupling strength is $G=0.2\omega_m$, the cavity decay is $\kappa=0.03\omega_m$, and the mechanical damping is $\gamma=10^{-3}\omega_m$. The stroke times are $\tau_1 = \tau_3 = 25\omega_m^{-1}$, $\tau_2 = 50\omega_m^{-1}$, and $\tau_4 = 10^4\omega_m^{-1}$. All the parameters are chosen according to the hierarchy relationship (\ref{eq:hierarchy}).} 
\label{fig:evolution} 
\end{figure*}

To simulate all aspects of the proposed Otto cycle, including nonadiabatic transitions between the lower and upper polariton branches and the effects of dissipation, we have solved numerically the full master equation (\ref{eq:master}). As already discussed, for small optomechanical couplings it is appropriate to start from the factorized thermal state
\begin{equation}
\rho_{{\rm sys}}(0)=\rho_{{\rm th}}^{a}\otimes\rho_{{\rm th}}^{b}
\end{equation}
where $\rho_{{\rm th}}^{a}$ and $\rho_{{\rm th}}^{a}$  are the thermal state of the photon and the phonon modes, respectively. 

Our numerical simulations were carried out in a Fock states basis of the bare modes, with a cutoff number state $|N\rangle$  with $N \gg \bar{n}_{a(b)}$, so that the total dimension of the density matrix $\rho_{\rm sys}$ is $(N+1)^4$. As a result the simulations become very time consuming even for relatively modest values of $N$. However, due to the diagonality of thermal states in an energy basis the total density matrix of the combined system remains in practice quite sparse for weak optomechanical couplings, and an algorithm optimization for sparse matrices is helpful to reduce the simulation time. To utilize the specific advantage of MATLAB in the calculation of large sparse matrices, we expressed all creation and annihilation operators in matrix form so that the master equation (\ref{eq:master}) was converted into a matrix equation, and then used a fourth-order Runge-Kutta algorithm to evolve the matrix equation, with convergence tested by increasing $N$.    

Figure~\ref{fig:evolution} shows the dynamics of the population of the bare modes $a$ and $b$ and of the polariton modes $A$ and $B$ for a full loop of the Otto cycle. We assume a small thermal mean phonon number $\bar{n}_{b}=4$ and truncate the number state representation at $N=30$. For a mechanical resonator of frequency $\omega_m=2\pi\times200\ \rm{MHz}$, it corresponds to a phonon reservoir temperature $T_b=45\ \rm{mK}$, which is in the parameter range of present optomechanical experiments. The mean photon number is zero. The initial detuning is $\delta_{i}=-3$ and we choose the final detuning $\delta_{f}=-0.4$ to avoid the unstable region near the cavity resonance. Other parameters are listed in the figure caption.

As shown in the figure, in the first stroke the population of the polaritons $A$ and $B$ initially coincides with the mean photon and phonon number, respectively. During the change in detuning, in the ideal case, the photon and the phonon mode would exchange their population while keeping the polariton numbers constant. In practice, the variation of the detuning is not slow enough to avoid transitions between the two polariton branches. Meanwhile, the cavity and mechanical damping rates result in a small decay and increment of the polariton $B$ and $A$, respectively. In the simulation we vary the detuning linearly in time, but an optimization of the pulse shape might result in a more effective transformation. For instance, one might change the detuning fast for large values and only slow down close to the avoided crossing at $\delta=-1$. 

In the second stroke, the photonlike polariton $B$ decays fast due to the cavity decay, while the thermalization of the phononlike polariton $A$, at rate $\gamma$, is negligible. This step is also characterized by Rabi oscillations  between photon and phonon populations due to the optomechanical coupling. The polariton $B$ then recovers its phononlike properties in the third stroke and finally rethermalizes to its initial population at the end of the last stroke (not shown, at $t>10^4\omega_{m}^{-1}$). The total cycle takes a time of the order $10^{-4}\ \rm s$. Note that the population of polariton $A$ remains small throughout the whole cycle, its effect on the engine can be safely neglected as we did in the previous discussion.

\begin{figure}
\includegraphics[width=0.4 \textwidth]{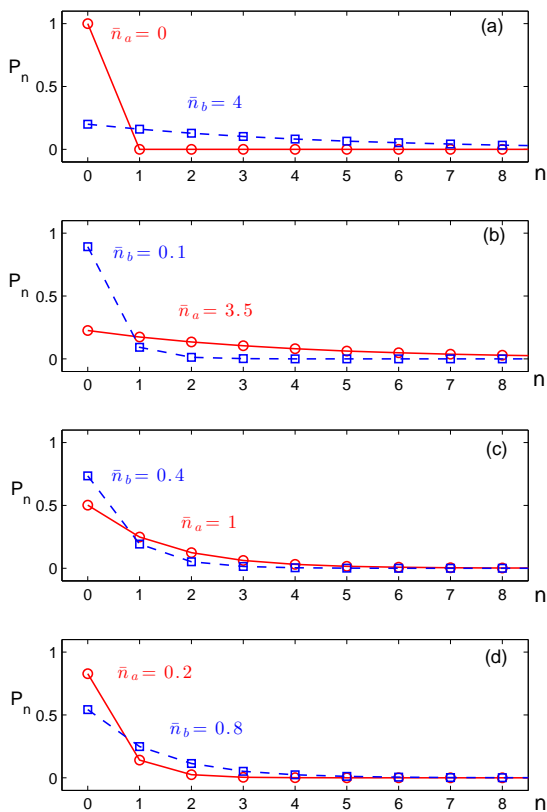}
\caption{(Color online)  The number state probability distribution $P_n$ of the quantum states of the photon mode $a$ (red solid line and circles) and the phonon mode $b$ (blue dashed line and squares). (a-d) correspond to the nodes of the Otto cycle (see text). $\bar{n}_{a(b)}$ are the corresponding mean particle numbers.} 
\label{fig:distribution} 
\end{figure}

Figure~\ref{fig:distribution} shows  the number state distribution of the photon and phonon modes at each node of the Otto cycle. The probabilities $P_n$ of the photon and phonon modes remain reasonably consistent with thermal-like distributions during the whole Otto cycle, but with a variable mean particle number $\bar{n}_{a}$ and $\bar{n}_{b}$. Comparing the initial [Fig.~\ref{fig:distribution}(a)] and final [Fig.~\ref{fig:distribution}(b)] situations for the first stroke,  the phonon and the photon almost exactly exchange their number distributions, except for a small particle loss. Initially both the photon and the phonon excitations are in their own thermal equilibrium with $\bar{n}_{a}$ and $\bar{n}_{b}$ determined by their reservoir temperatures, but after following the adiabatic change in detuning they are in nonequilibrium states, with no well-defined temperatures. At this point both $\rho_a$ and $\rho_b$ have nonvanishing nondiagonal elements, as the density matrix $\rho_{ab}$ of the full system exhibits quantum correlation between the photon and the phonon resulting from the optomechanical interaction~(\ref{V}). 
They tend to return to thermal equilibrium during the second stroke [from Fig. \ref{fig:distribution}(b) to Fig. \ref{fig:distribution}(c)], but the phonon mode is prevented from completely doing so, given the short time compared to its dissipation rate. As we recall from the previous section this is a necessary requirement to guarantee that the polariton $A$ population remains small. A conversion similar to the first one happens in the third stroke, [from Fig. \ref{fig:distribution}(c) to Fig. \ref{fig:distribution}(d)]. Finally the photon and the phonon recover their initial thermal equilibrium [Fig  \ref{fig:distribution}(a)] after a time that is long compared to $\gamma^{-1}$.

\section{Summary and outlook}
In summary, we have presented a detailed theoretical study of a simple quantum optomechanical heat engine based on changing the nature of its normal modes from phononlike to photonlike, and we provided an intuitive picture of its operation. We have discussed the performance of the engine, recovering a quantum version of the Curzon-Ahlborn efficiency at maximum work. Much insight into the underlying physics of this system can be gained by comparing its description in terms of normal modes (polaritons) and bare modes. We found that the polariton description provides an unambiguous distinction between heat exchange and work, but there are subtleties associated with the fact that polaritons are generally coupled to squeezed reservoirs when the bare modes are coupled to thermal reservoirs. In the case of strong optomechanical coupling this may lead to novel features, including possibly a violation of the Carnot efficiency limit, which will be explored in future work. The bare mode picture is intuitively appealing and illustrates clearly the role of quantum correlations between photons and phonons in the work acting on the system but lacks a clear decomposition of the Otto cycle into adiabatic and thermalization strokes.  Numerical simulations of the full quantum system coupled to thermal baths confirm, however, the general intuition of the engine operation in the weak optomechanical coupling limit. 

Future work will consider schemes to measure and exploit the work performed by the engine, including the effect of quantum back-action and its possible impact on the efficiency of the engine. We will also explore in detail the roles of entanglement and polariton reservoir squeezing on the Otto cycle.

\section{Acknowledgments}
We thank H. Seok for helpful discussions. This work was supported by the National Basic Research Program of China under Grant No. 2011CB921604, the NSFC under Grants No.~11204084 and 11234003, the Specialized Research Fund for the Doctoral Program of Higher Education No.~20120076120003, the SCST under Grant No.~12ZR1443400, the DARPA QuASAR and ORCHID programs through grants from AFOSR and ARO, the U.S. Army Research Office, and the US NSF.

\end{document}